\documentstyle[bezier,epsfig,12pt,preprint,tighten,aps]{revtex}
\begin{document}

\draft

\title{\rightline{{\tt (February 1998, revised April 1998)}}
\rightline{{\tt TMUP-HEL-9803}}
\rightline{{\tt UM-P-98/06}}
\rightline{{\tt RCHEP-98/02}}
\ \\
Confronting solutions to the atmospheric neutrino anomaly\\
involving large angle $\nu_{\mu} \to \nu_e$ oscillations with\\
SuperKamiokande and CHOOZ}
\author{R. Foot and R. R. Volkas}
\address{School of Physics\\
Research Centre for High Energy Physics\\
The University of Melbourne\\
Parkville 3052 Australia\\
(foot@physics.unimelb.edu.au, r.volkas@physics.unimelb.edu.au)}
\author{O. Yasuda}
\address{Department of Physics\\
Tokyo Metropolitan University\\
1-1 Minami-Osawa Hachioji, Tokyo 192-0397, Japan\\
(yasuda@phys.metro-u.ac.jp)}
\maketitle

\begin{abstract} 

Neutrino oscillation scenarios involving large
angle $\nu_{\mu} \to \nu_e$ oscillations are disfavoured in the
parameter range $\Delta m^2/eV^2 \stackrel{>}{\sim} 10^{-3}$
by recent results from the CHOOZ reactor-based
$\overline{\nu}_e$ disappearance experiment.  
For this reason we extend our previous work on up-down 
asymmetries for various oscillation scenarios by computing 
up-down asymmetries and the R ratio for the
entire conceivable range $10^{-4} - 10^{-1}\ eV^2$ of $\Delta m^2$. 
Matter effects in the Earth play a crucial role. 
We perform a $\chi^2$ fit to the
data. We find that, because of the matter effect, 
the three-flavour maximal mixing model provides a reasonable 
fit to SuperKamiokande and CHOOZ
data provided that the relevant $\Delta m^2$ is in the range
$4 \times 10^{-4} \stackrel{<}{\sim} 
\Delta m^2/eV^2 \stackrel{<}{\sim} 1.5 \times 10^{-3}$. 

\end{abstract}

\newpage

Recent data from atmospheric neutrino experiments \cite{tot}
and especially the SuperKamiokande experiment \cite{tot,sk}
provide very strong evidence for large angle
neutrino oscillations.
Traditionally the atmospheric neutrino anomaly has 
been represented by the quantity R, where
\begin{equation} 
R \equiv \frac{(N_{\mu}/N_e)|_{data}}{(N_{\mu}/N_e)|_{MC}}.  
\end{equation}
The quantities $N_{e,\mu}$ are the numbers of $e$-like and 
$\mu$-like events.
In addition to an anomalous value for $R$, the Kamiokande and
SuperKamiokande experiments have observed anomalous
zenith angle dependence \cite{sk,fuk}.
This zenith angle dependence can be represented by
the up-down asymmetry parameters \cite{bfv,flp,fvy,fvy2,flmm} 
\begin{equation}
Y^{\eta}_{\alpha} \equiv 
{(N_{\alpha}^{-\eta}/N_{\alpha}^{+\eta})|_{data}
\over (N_{\alpha}^{-\eta}/N_{\alpha}^{+\eta})|_{MC}}\ \ 
(\alpha = e,\mu).
\end{equation}
Here $N_{\alpha}^{-\eta}$ denotes the number 
of $\alpha$-like events produced in
the detector with zenith angle $\cos \Theta < -\eta$, while
$N_{\alpha}^{+\eta}$ denotes the analogous quantity for $\cos \Theta >
\eta$, where $\eta$ is defined to be positive (note that
$\cos \Theta > 0$ for downward going leptons). 
SuperKamiokande divides the
$(-1,+1)$ interval in $\cos\Theta$ into five equal bins. 
The central bin straddles both the upper and lower 
hemispheres, and is thus not useful for
up-down asymmetry analyses. We therefore choose $\eta = 0.2$ 
in order to utilise all the data in the other four bins. 

When comparing the measured results for $R$ and the 
$Y$'s with predictions from a specific neutrino oscillation 
model, the numerators are replaced by
calculated predictions from the models, while the 
denominators remain as the no-oscillation predictions. 
Note that systematic uncertainties for
up-down asymmetries are expected to be smaller than 
for $R$, because the latter depends on the relative flux of 
$\nu_{\mu}$ to $\nu_e$. 

The utility of using up-down asymmetries as a 
diagnostic tool was emphasised in Ref.\cite{flp}, 
where up-down asymmetries were computed for
various neutrino oscillation solutions to the 
atmospheric neutrino anomaly.
The analysis of Ref.\cite{flp} focussed on the 
energy dependence of up-down
asymmetries. By contrast, in Refs.\cite{fvy,fvy2} we 
followed (super)Kamiokande and considered particular 
cuts in energy. 

In Ref.\cite{fvy2}, we analysed four representative
cases:
\vspace{1mm}

\noindent
(A) Maximal $\nu_{\mu} - \nu_{\tau}$ 
mixing \cite{barpak}.

\vspace{1mm}

\noindent
(B) Maximal $\nu_{\mu} - \nu_e$ mixing\cite{ap}.

\vspace{1mm}

\noindent 
(C) Threefold maximal mixing \cite{giunti,hps} amongst $\nu_e$,
$\nu_{\mu}$ and $\nu_{\tau}$.

\vspace{1mm}

\noindent
(D) Massless neutrinos with violation of the Equivalence Principle or
breakdown of Lorentz invariance \cite{gasp}. The case of 
maximal $\nu_e - \nu_{\mu}$ oscillations\cite{fnn}
was considered for definiteness.

\vspace{1mm}

In Ref.\cite{fvy2}, we focussed on the region of parameter space $\Delta
m^2/eV^2 > 2\times 10^{-3}$. However, large angle $\nu_{\mu} \to \nu_e$
oscillations are now disfavoured in this parameter range because of recent
results from the CHOOZ reactor-based $\overline{\nu}_e$ disappearance
experiment\cite{chooz}. This experiment disfavours maximal $\nu_e -
\nu_{\mu}$ mixing for $\Delta m^2/eV^2 > 0.9 \times 
10^{-3}$ at 90\% C.L.  Thus 
it is important to discuss the parameter space region $\Delta
m^2/eV^2 < 2\times 10^{-3}$.  In Ref\cite{fvy2} we neglected matter effects
\cite{msw} due to neutrino oscillations through the Earth.  However, it
turns out that this is {\it not} a valid approximation for the
2-flavour cases for multi-GeV (sub-GeV)
neutrinos unless $\Delta m^2/eV^2 \stackrel{>}{\sim} 10^{-2}$
($10^{-3}$), and
is never a good approximation for model C. 
Thus the purpose of this
paper is to reexamine the up-down asymmetries and $R$ for the entire
conceivable range of interest for $\Delta m^2$ (i.e. $\Delta m^2/eV^2 >
10^{-4}$).  We will numerically integrate
the Schr\"odinger equation for neutrino evolution including the matter 
effects, taking the density profile of the Earth from Ref.\cite{stacy}.  
We will show that for the two flavour $\nu_{\mu} \to \nu_e$ oscillation
models B and D the matter effects suppress the oscillations and do not
improve the fit of these models to the data.  {\it Interestingly, however,
for the 3-flavour model C the matter effects actually improve the fit of
the model to the data.} Although this model does not
fit the data as well as model A, we will show that
this model does provide an acceptable fit to the 
SuperKamiokande and CHOOZ data for a range
of $\Delta m^2$.

Our methodology is similar to our previous paper \cite{fvy2} except that we
have used the inclusive cross section given in Ref.\cite{bp} for the
multi-GeV analysis.  Although this cross section is not completely
satisfactory for calculating absolute event rates because it does not
incorporate low $Q^2$ effects such as the $\Delta$ resonance production, it
is a good enough approximation for calculating ratios of event rates such
as up-down asymmetries and $R$. We also include results for case A, even
though it does not involve $\nu_e$, because it will be interesting to
compare cases A and C. [For a comparative analysis of case A and the
somewhat similar large angle $\nu_{\mu} \to \nu_s$ solution (where $\nu_s$
is a sterile neutrino), see Ref,\cite{fvy3}].  We are now able to improve
on the
analysis of Ref.\cite{fvy2} in another respect, because we have been
fortunate to obtain the detection efficiency functions from the
SuperKamiokande collaboration. 

Our results are given in Figures 1-6 \cite{fn5,fnx}, together with 
preliminary results from superKamiokande (corresponding
to 414 days of live running)\cite{sk} 
\begin{eqnarray}
R ({\it sub-GeV}) & = & 0.61 \pm 0.03 \pm 0.05,\ 
R ({\it multi-GeV}) = 0.67 \pm 0.06 \pm 0.08, \nonumber\\
Y^{0.2}_{\mu} ({\it sub-GeV}) & = & 0.78 \pm 0.06,\ 
Y^{0.2}_{\mu} ({\it multi-GeV}) = 0.49 \pm 0.06, \nonumber\\ 
Y^{0.2}_{e} ({\it sub-GeV}) & = & 1.13 \pm 0.08,\
Y^{0.2}_{e} ({\it multi-GeV}) = 0.83 \pm 0.13.
\end{eqnarray}
Note that only statistical errors are given for the up-down 
asymmetries since they should be much larger 
than possible systematic errors at the
moment.

Figures 1,2 show that all of the models A,B,C,D can provide
an acceptable fit to $R$. However the up-down
asymmetries $Y_{e,\mu}$ clearly distinguish the models.
The only cases which can provide
an acceptable fit to all of the data are A and C.
Indeed, the $Y_{e,\mu}$ and $R$ values for model C
are quite similar to model A for low $\Delta m^2$.
To understand this point, consider the
Schr\"odinger equation for neutrino evolution 
in model C
including matter effects,
\begin{equation}
i {d \over dx} \left[ \begin{array}{c} \nu_e (x) \\ \nu_{\mu}(x) \\ 
\nu_{\tau}(x)
\end{array} \right] = 
{\Delta m^2 \over 2E}
\left[ \begin{array}{ccc}
A(x) + 1/3  & 1/3 & 1/3 \\
1/3  & 1/3 & 1/3 \\
1/3  & 1/3 & 1/3 
\end{array} \right]
\left[ \begin{array}{c} \nu_e (x) \\
\nu_{\mu}(x) \\ \nu_{\tau}(x)
\end{array} \right],
\end{equation}
where $x$ is the distance travelled, $E$ the neutrino energy, $\Delta m^2$
the larger of the two squared-mass differences in model C
and $\nu_{e,\mu,\tau}(x)$ the
wave-functions of the neutrinos. The quantity $A(x)$
is related to the effective potential difference 
generated through the matter effect:
\begin{equation}
A(x) = {2E \over \Delta m^2}\sqrt{2} G_F N_e(x) 
\simeq 2.9\times 10^{-4}.
\left[{E/GeV \over \Delta m^2/eV^2}\right]
\left[ {\rho (x) \over 4\ g/cm^3}
\right],
\label{mon}
\end{equation}
where $G_F$ is the Fermi constant, $N_e(x)$ the number density of
electrons along the path of the neutrino 
and $\rho(x)$ the mass density of the earth (in deriving
the right hand side of Eq.(\ref{mon}) we have assumed
that the average number of protons per nucleon is approximately
constant $\sim 0.48$).
For antineutrinos the sign of $A(x)$ is reversed.

For $E/\Delta m^2$ values leading to large matter effects ($A \gg
1/3$),  $\nu_e$ oscillations are suppressed. The system in this case
exhibits approximate two flavour $\nu_\mu \to \nu_\tau$
maximal oscillations with
\begin{equation}
P(\nu_\mu \to \nu_\tau) = 1 - \sin^2 \left[{2 \over 3}
\times 1.27{(L/km)(\Delta m^2/eV^2) \over (E/GeV)}\right].
\label{zz}
\end{equation}
This qualitatively explains why $R,Y$ for case C are
similar to case A for $\Delta m^2/eV^2 \stackrel{>}{\sim}
3 \times 10^{-3}$.  From Eq.(\ref{zz}), observe that 
for a given $\Delta m^2$ 
the oscillation length of $\nu_\mu \to \nu_\tau$ 
is not the same as the oscillation length for genuine two 
flavour oscillations - it is $1.5$ times longer.
This explains why the multi-GeV $R$ and $Y_{\mu}$ 
for model C are displaced relative to model A.
Finally note that in the large $\Delta m^2/E$ limit, the
$Y_\mu$ asymmetry does not approach 1 in model C.
In fact, one can show by explicit computation that 
in the large $\Delta m^2/E$ limit, the muon neutrino
survival probability is related to the vacuum survival
probability by
\begin{equation}
P(\nu_\mu \to \nu_\mu)(A) = P(\nu_\mu \to \nu_\mu)(A=0) - {1 \over 6}
\left(1 - \cos {2 \over 3}{\Delta m^2AL \over 2E}\right).
\end{equation}
Note that $\Delta m^2 A/E$ is independent of $\Delta m^2$ and $E$ 
and thus it turns out that there is no range of parameters where 
matter effects can be neglected for atmospheric neutrinos in
model C.

We now perform a $\chi^2$ analysis to determine the preferred region of
$\Delta m^2$ for model C.  We will not
consider models B and D because they
obviously lead to bad fits.
We first define a $\chi^2$ function for atmospheric data:
\cite{gau}
\begin{equation}
\chi^2_{atm} = \sum_E \left[\left({R^{SK} - R^{th} \over \delta
R^{SK}}\right)^2
+ \left({Y^{SK}_{\mu} - Y^{th}_{\mu} \over \delta Y^{SK}_{\mu}}\right)^2
+ \left({Y^{SK}_{e} - Y^{th}_{e} \over \delta Y^{SK}_{e}}\right)^2
\right],
\end{equation}
where the sum is over the sub-GeV and multi-GeV cases, the measured
SuperKamiokande values and errors are denoted by the superscript ``SK''
and the theoretical predictions for the quantities are labelled by ``th''.
The $\eta = 0.2$ choice is understood for the up-down asymmetries. 
There are 6 pieces of data 
in $\chi^2$ and 1 adjustable parameter,
$\Delta m^2$, leaving 5 degrees of freedom. [Note that in
the present paper we consider 
$|U_{\alpha j}|^2 = 1/3~(\alpha = e, \mu, \tau, \ j=1,2,3$)
for model C and thus the mixing angles do
not constitute free parameters]. 

The solid line in Fig. 7 displays $\chi^2_{atm}$ as a function of $\Delta
m^2$ for model C (also shown is $\chi^2_{atm}$ for the $Y$ asymmetries
only). The CHOOZ experiment disfavours $\Delta m^2/eV^2
\stackrel{>}{\sim} 10^{-3}$ for large angle 
$\nu_{\mu} \to \nu_e$ oscillations\cite{chooz}.
In order to incorporate
the CHOOZ results, we define another $\chi^2$:
\begin{equation}
\chi^2_{CHOOZ} = 
\sum_i \left( {x_i - y_i \over \delta x_i} \right)^2.
\end{equation}
The sum is over 12 energy bins of data (in the
above equation $x_i$ are experimental values from
Figure 5b of Ref.\cite{chooz} and $y_i$ are the 
corresponding theoretical predictions)\cite{my}. 
The short-dashed line in Fig.7 displays
$\chi^2_{CHOOZ}$ as a function of $\Delta m^2$ for model C. Figure 
8 plots $\chi^2_{atm}+\chi^2_{CHOOZ}$. The $3\sigma$ allowed range for the
atmospheric data plus CHOOZ is
\begin{equation}
4 \times 10^{-4} \stackrel{<}{\sim} \Delta m^2/eV^2 \stackrel{<}{\sim} 1.5
\times 10^{-3}.
\end{equation}
The best fit point at $\Delta m^2 \simeq 8 \times 10^{-4}\ eV^2$ gives
$\chi^2_{min} \simeq 23$ for 17 degrees of freedom 
which implies an allowed C.L. of $16\%$ 
for model C to explain the atmospheric
data while simultaneously being consistent with CHOOZ.

We have also performed a $\chi^2$ fit for model C to the solar neutrino
data for the parameter space of Ref.\cite{hps}. This
model predicts an energy independent 
solar electron neutrino flux reduction of $5/9$ (this
leads to predictions for the solar neutrino experiments which
are reduced by a factor $5/9$, except for (super)Kamiokande where
neutral current effects must be incorporated).
We have used the most recent results for Homestake ($2.55
\pm 0.14 \pm 0.14 \ SNU$)\cite{hc}, 
GALLEX($76.4 \pm 6.3^{+4.5}_{-4.9}\ SNU$)\cite{x1}, 
SAGE ($69.9^{+8.9}_{-8.7}\ SNU$)\cite{x2}.
For Kamiokande (SuperKamiokande) we have used the 8(16) energy 
bins given in Ref.\cite{x3} (Ref.\cite{x4}).
We choose to leave the boron flux as a free parameter\cite{ap,hps} and
neglect the small theoretical errors of the other fluxes.
We consider two solar models for definiteness,
BP95\cite{BP95} and TCL\cite{TCL}.
We find that $\chi^2_{solar} (min) = 44$ for BP95 solar model
and $\chi^2_{solar} (min) = 33$ for TCL solar model.
With the boron flux as a free-parameter, there are $27-1 = 26$ 
degrees of freedom.
Thus, the overall fit of model C to the solar+atmospheric+CHOOZ
experiments turns out to be reasonable with
$\chi^2_{min} = 67$  (BP95)  and  $56$  (TCL) for $43$ 
degrees of freedom which corresponds to an allowed
C.L. of $1\%$ (BP95) and $10\%$ (TCL).

In conclusion, we have extended the analysis of Ref.\cite{fvy2} 
to include low values of $\Delta m^2/eV^2$.
This is important for models which have large angle $\nu_{\mu} \to \nu_e$ 
oscillations because of the recent CHOOZ results. Out of the three cases
(B, C and D) which involve $\nu_e$, only the 
three-flavour maximal mixing model (case C) provides a reasonable fit to 
the up-down asymmetries and $R$ ratios while being consistent with CHOOZ. 
We thus reach the important conclusion that the most favoured solutions to
the atmospheric neutrino anomaly in the light of CHOOZ are: (i) large angle
or maximal 
$\nu_{\mu} \to \nu_{\tau}$ oscillations (case A), (ii) large angle or
maximal $\nu_{\mu}
\to \nu_s$ oscillations \cite{fvy3} and, (iii) three-flavour 
maximal mixing
(case C) with $\Delta m^2 \sim 8 \times 10^{-4}\ eV^2$ (sufficiently small 
departures from three-flavour maximal mixing would also of course be
allowed). If (i) is true then
the planned Japanese long baseline neutrino oscillation 
experiment may or may not see a
signal (see Ref.\cite{fvy3}), if (ii) is true then present data suggest
that they
should see a signal \cite{fvy3}, and if three-flavour maximal mixing is
correct then the Japanese long baseline experiment 
has no chance to see a signal.
Finally note that if model C is correct then the Kamland experiment 
should see a positive signal.

\acknowledgments{O.Y.
was supported in part by a Grant-in-Aid for Scientific Research of the
Ministry of Education, Science and Culture, \#09045036. O.Y. would like to
thank T. Kajita for useful communications and the participants of the
Neutrino Symposium at Hachimantai, Japan, on Nov.28-30 1997 for
discussions. R.F. and R.R.V. are supported by the Australian Research
Council. R.F. would like to thank P. Harrison for useful correspondence.

\newpage

\centerline{\large \bf Figure Captions}

\vspace{5mm}

\noindent
Figure 1.\ \ The sub-GeV $R$ as a function of $\Delta m^2/eV^2$ 
[$(\delta v/2)km GeV$] for models A (solid line), B (dashed-dotted
line), C (dashed line) [D (dotted line)].
The usual SuperKamiokande sub-GeV momentum cuts have 
been employed. 
The horizontal long-dashed lines are the preliminary SuperKamiokande 
data $\pm 1 \sigma$ statistical errors.

\vspace{5mm}

\noindent 
Figure 2.\ \ Same as Figure 1 except for SuperKamiokande 
multi-GeV case.

\vspace{5mm}

\noindent
Figure 3.\ \ The sub-GeV up-down $e$-type asymmetry $Y^{0.2}_{e}$
as a function of $\Delta m^2/eV^2$ 
[$(\delta v/2)km GeV$] for models A (solid line), B (dashed-dotted
line), C (dashed line) [D (dotted line)].
The usual SuperKamiokande sub-GeV momentum cuts have 
been employed. 
The horizontal long-dashed lines are the preliminary SuperKamiokande 
data $\pm 1 \sigma$ statistical errors.

\vspace{5mm}

\noindent
Figure 4.\ \ Same as Figure 3 except for muons instead of electrons.

\vspace{5mm}

\noindent 
Figure 5.\ \ Same as Figure 3 except for multi-GeV sample.

\vspace{5mm}

\noindent
Figure 6.\ \ Same as Figure 3 except for muons and 
for multi-GeV sample.

\vspace{5mm}
\noindent
Figure 7.\ \ $\chi^2$ fit as a function of $\Delta m^2$ to the
SuperKamiokande atmospheric data for Model C.  The solid line
includes both $R$ and up-down asymmetries whereas the dashed
line includes only the up-down asymmetries.
The $\chi^2$ for the CHOOZ reactor data (dotted line) is also shown.

\vspace{5mm}

\noindent
Figure 8.\ \ Combined $\chi^2$  fit as a 
function of $\Delta m^2$ to the SuperKamiokande atmospheric 
data and the CHOOZ reactor data for model C.

\newpage
\epsfig{file=f1.eps,width=15cm}
\newpage
\epsfig{file=f2.eps,width=15cm}
\newpage
\epsfig{file=f3.eps,width=15cm}
\newpage
\epsfig{file=f4.eps,width=15cm}
\newpage
\epsfig{file=f5.eps,width=15cm}
\newpage
\epsfig{file=f6.eps,width=15cm}
\newpage
\epsfig{file=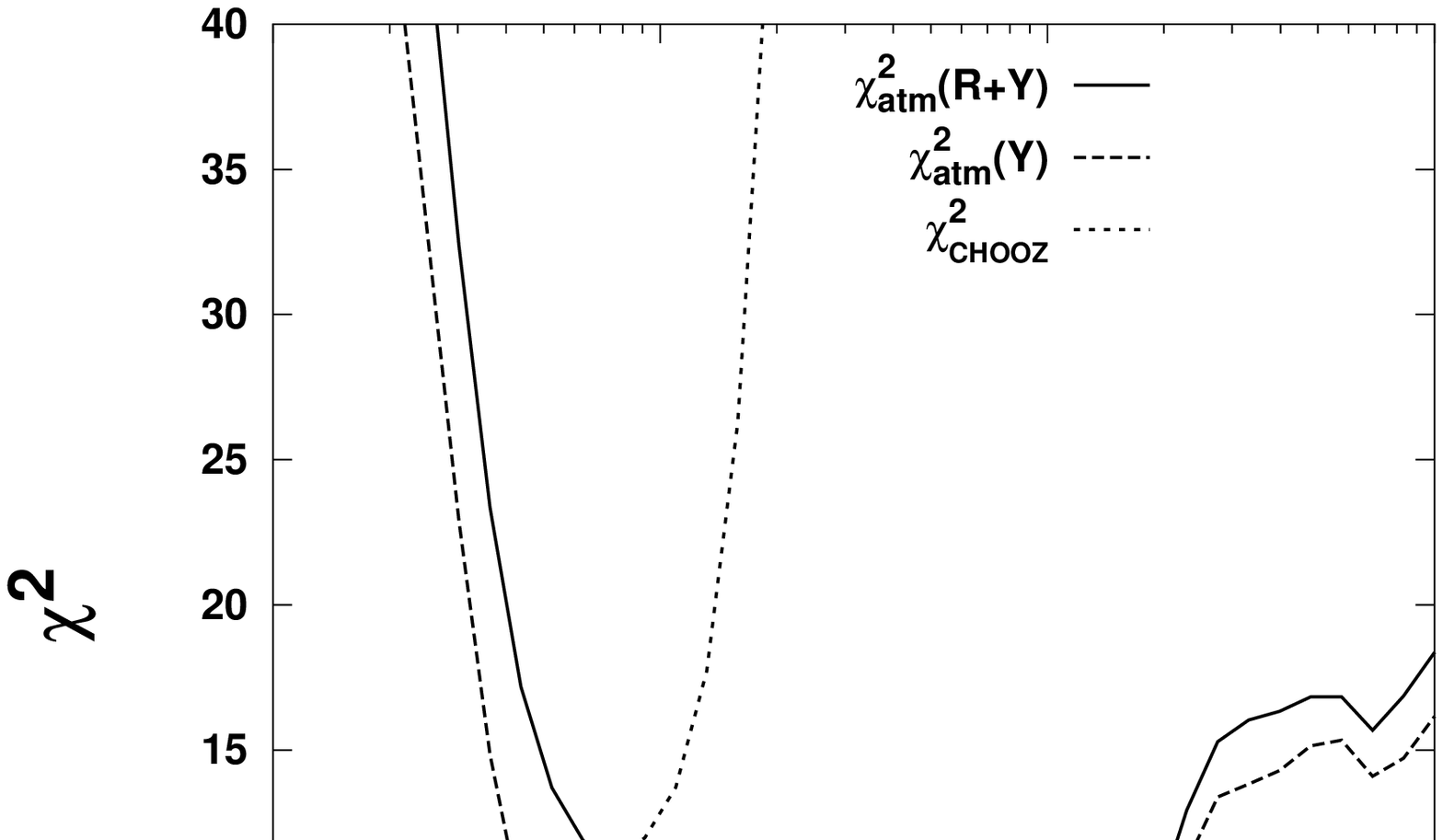,width=15cm}
\newpage
\epsfig{file=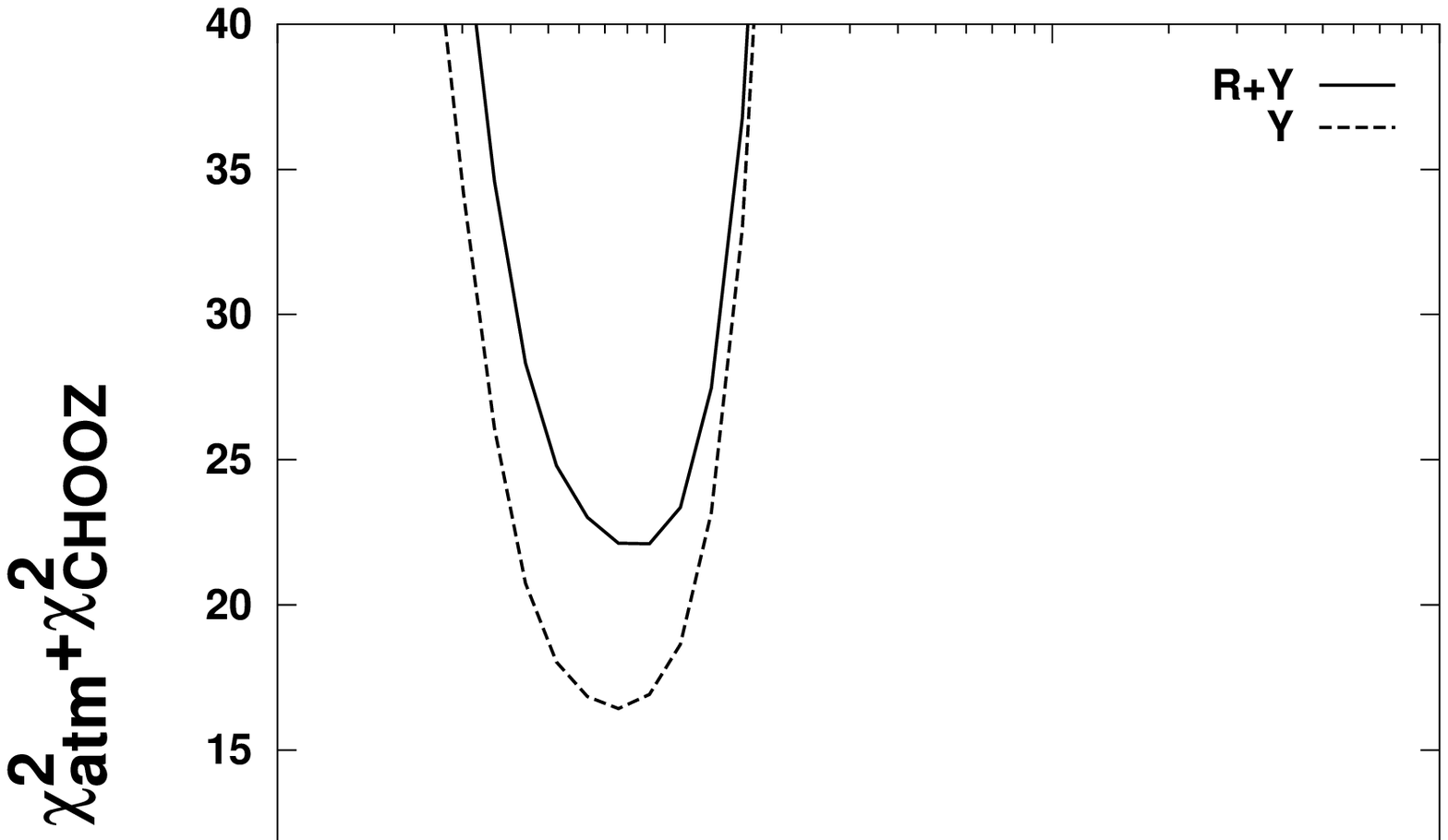,width=15cm}

\begin{thebibliography}{99}
\bibitem{fnn}
In this case the survival probability is given by
$P_{ee} = P_{\mu \mu} = 1 - \sin^2 {\delta v \over 2} E L$
and the transition probability is $P_{e\mu} = P_{\mu e} 
= \sin^2 {\delta v \over 2} E L$ (where $E$ and $L$ are the
energy of the neutrino and the distance travelled respectively).
\bibitem{fn5}
Note that we have included the effects of 
an estimated $7\%$\cite{sk}
misidentification of muons as electrons in the multi-GeV analysis.
For this reason $\nu_{\mu} \to \nu_{\tau}$ oscillations
can slightly affect the $Y_e$ asymmetries.
\bibitem{fnx}
In our numerical work we found some small oscillations (less
than a few percent) which we have smoothed out in the figures.
\end{thebibliography}
\end{document}